\documentclass[twocolumn]{aastex631}
\usepackage{amsmath,amstext}
\usepackage[T1]{fontenc}
\usepackage{apjfonts}
\usepackage{graphicx}
\usepackage{placeins}
\usepackage{xspace}

\newcommand{\firstpaper}{E19}
\newcommand{\Aeos}{A{\sc eos}\xspace}

\usepackage{placeins}

\newcommand{\HI}{{H\sc~i}}

\begin{document}

\title{\textsc{Aeos}: Star-by-Star Cosmological Simulations of Early Chemical Enrichment and Galaxy Formation}

\correspondingauthor{Kaley Brauer}
\email{kaley.brauer@cfa.harvard.edu}


\author[0000-0002-8810-858X]{Kaley Brauer}
\affiliation{Center for Astrophysics | Harvard \& Smithsonian, Cambridge, MA 02138, USA}

\author[0000-0003-2807-328X]{Andrew Emerick}
\affiliation{Carnegie Observatories, Pasadena, CA, 91101, USA}

\author[0009-0006-4744-2350]{Jennifer Mead}
\affiliation{Department of Astronomy, Columbia University, New York, NY 10027, USA}

\author[0000-0002-4863-8842]{Alexander P. Ji}
\affiliation{Department of Astronomy \& Astrophysics, University of Chicago, 5640 S Ellis Avenue, Chicago, IL 60637, USA}
\affiliation{Kavli Institute for Cosmological Physics, University of Chicago, Chicago, IL 60637, USA}

\author[0000-0003-1173-8847]{John H. Wise}
\affiliation{Center for Relativistic Astrophysics, School of Physics, Georgia Institute of Technology, Atlanta, GA 30332, USA}

\author[0000-0003-2630-9228]{Greg L. Bryan}
\affiliation{Department of Astronomy, Columbia University, New York, NY 10027, USA}
\affiliation{Center for Computational Astrophysics, Flatiron Institute, New York, NY 10010, USA}

\author[0000-0003-0064-4060]{Mordecai-Mark Mac Low}
\affiliation{Department of Astrophysics, American Museum of Natural History, New York, NY 10024, USA}
\affiliation{Department of Astronomy, Columbia University, New York, NY 10027, USA}

\author[0000-0002-9986-8816]{Benoit C{\^o}t{\'e}}
\affiliation{Department of Physics and Astronomy, University of Victoria, Victoria, BC, V8P5C2, Canada}

\author[0000-0003-3479-4606]{Eric P. Andersson}
\affiliation{Department of Astrophysics, American Museum of Natural History, New York, NY 10024, USA}

\author[0000-0002-2139-7145]{Anna Frebel}
\affiliation{Department of Physics and Kavli Institute for Astrophysics and Space Research, Massachusetts Institute of Technology, Cambridge, MA 02139, USA}

\keywords{Galaxy chemical evolution -- Dwarf galaxies -- Chemical enrichment -- Hydrodynamics}

\begin{abstract}

The \Aeos project introduces a series of high-resolution cosmological simulations that model star-by-star chemical enrichment and galaxy formation in the early Universe, achieving 1 pc resolution. These simulations capture the complexities of galaxy evolution within the first $\sim$300~Myr by modeling individual stars and their feedback processes. By incorporating chemical yields from individual stars, \Aeos generates galaxies with diverse stellar chemical abundances, linking them to hierarchical galaxy formation and early nucleosynthetic events. These simulations underscore the importance of chemical abundance patterns in ancient stars as vital probes of early nucleosynthesis, star formation histories, and galaxy formation. We examine the metallicity floors of various elements resulting from Pop III enrichment, providing best-fit values for eight different metals (e.g., [O/H] = -4.0) to guide simulations without Pop III models. Additionally, we identify galaxies that begin star formation with Pop II after external enrichment and investigate the frequency of CEMP stars at varying metallicities. The \Aeos simulations offer detailed insights into the relationship between star formation, feedback, and chemical enrichment. Future work will extend these simulations to later epochs to interpret the diverse stellar populations of the Milky Way and its satellites.
\end{abstract}

\section{Introduction}

The first galaxies formed 200--300 million years after the Big Bang, hosting many of the first stars and seeding the creation of every galaxy in the Universe today \citep{Bromm11}. The smallest of these early galaxies formed stars for 1--2 Gyr before being quenched by cosmic reionization and stellar feedback \citep{Brown14}, so those that survived until now are composed of ancient stars from 13 Gyr ago. Stars from these {\em ultra-faint dwarf galaxies} ($L<10^5 L_\odot$) are therefore relics from the era of the first stars and galaxies, preserving  signatures of early chemical enrichment \citep{2009ApJ...693.1859B, Simon2019}. 

Early galaxies played pivotal roles in the cosmic story. Over time, they merged to form the galaxies we observe today, including the Milky Way. Stars stripped from these dwarf galaxies likely comprise the metal-poor component of the Milky Way's stellar halo \citep{FrebelNorris2015}. These galaxies are also tied to reionization both as victims of quenching and as important sources of ionizing photons \citep{Brown14,Wise2014}. Due to their age and limited star formation histories, they offer invaluable insights into the earliest stages of star formation and chemical enrichment \citep[e.g.,][]{Ji2015}.
In the past twenty years, dozens of these ultra-faint dwarf galaxies have been discovered in the Local Group \citep[e.g.,][]{Willman2005,Bechtol15,Simon2019}. High-resolution spectroscopy has provided chemical abundances of over 15 of these dwarfs, with many more to come in the next decade. Simultaneously, chemical abundances of millions of stars in the Milky Way stellar halo are being obtained through wide-field spectroscopic programs such as GALAH (\citealt{DeSilva15}), H3 (\citealt{Conroy19}), APOGEE (\citealt{Majewski17}), RAVE (\citealt{Steinmetz06}), SEGUE (\citealt{Yanny09}), DESI (\citealt{Desi20}), and LAMOST (\citealt{Cui12}), with more to come from WEAVE (\citealt{Dalton14}),  4MOST (\citealt{deJong19}), SDSS-V (\citealt{Kollmeier17}), and the Rubin Observatory (\citealt{Rubin19}). 

A key pathway to understanding early star and galaxy formation is through the study of stellar chemical abundances in ancient stars, particularly those found in ultra-faint dwarfs and the Milky Way's stellar halo \citep[e.g.,][]{Brauer19}. These chemical abundances serve as ``fingerprints'', preserving crucial information about prior Pop III and Pop II star formation, hierarchical galaxy assembly, and nucleosynthetic processes. This is especially significant for stars that formed in the smallest, earliest galaxies, as their kinematic signatures are rapidly lost during mergers \citep[e.g.,][]{Brauer22}, but their chemical compositions remain detectable. However, the complexity of observed abundance patterns in dwarf galaxies has surpassed the capabilities of many current theoretical models of chemical evolution \citep[e.g.,][]{Ji20}, many of which still rely on the one-zone models of \citet{Tinsley80}.

Chemical evolution models that rely on simple parametric approaches and assume homogeneous mixing are capable of reproducing mean trends but incapable of modeling scatter \citep[e.g.,][]{Andrews17,Cote18}. When chemical abundance observations of dwarf galaxies were sparse, these models were sufficient to explain observed trends and differences between dwarf galaxies as a function of luminosity \citep[e.g.,][]{Tolstoy09,Kirby11}. Recent observations, however, have revealed abundance scatter in many elements both within and between dwarf galaxies \citep[e.g.,][]{Hill19,Ji20,Mead2024}. Such variations from mean trends imply the presence of complex galaxy formation processes like source- and time-dependent metal mixing, hierarchical galaxy merging, and bursty star formation \citep[e.g.,][]{Emerick2020a}. 

Over the last few decades, significant work has been done studying the chemodynamical evolution of galaxies using cosmological hydrodynamics simulations \citep[e.g.,][]{Oppenheimer08,Wiersma09,Shen10, Simpson13, Schaye10, Schaye15,Hopkins14, Agertz21, Renaud21, Buder24}. These simulations, coupled with increased attention to feedback processes, have made substantial strides in reproducing global galaxy trends, such as the evolution of the mass–metallicity relationship \citep[e.g.,][]{Obreja14,Ma16,Dave17}, and more detailed quantities like metallicity distribution functions and the evolution of individual species abundances \citep{Marcolini08, Revaz09, Sawala10, Revaz12, Jeon17, Hirai17}. Due to simple chemical enrichment models and resolution, however, these simulations are all unable to explain the observed chemical abundance scatter in Milky Way dwarf galaxies. 

More recently, hydrodynamic galaxy simulations have begun to reach parsec resolution and include star particles representing individual stars \citep[e.g.,][]{Emerick2019,Hirai21,Gutcke21,Hislop22,Andersson23,Lahen2020,Steinwandel23,Calura22,Deng24}. This level of detail allows simulations to capture variations in when, where, and how individual stars inject energy and metals. 
Such detail in a simulation allows investigation into the origins of the full chemical abundance distributions observed in dwarf galaxies beyond just the mean elemental trends.
\citet{Revaz2016} showed that below mass resolution of 10$^3$ M$_\odot$, stellar populations are not representative of the full IMF. This leads to oversimplifications in enrichment and the inability to properly model abundance scatter. This was similarly argued by \citet{Smith21} for mass resolution below 500 M$_\odot$, finding that stochastic variations in stellar populations become important when modeling radiation and galactic wind driving.
In the smallest galaxies, this becomes especially important, as each individual feedback event affects the evolution of the galaxy.
To understand the origin of these distributions and interpret both present and future chemical abundance data, simulations need to effectively combine extremely high-resolution star particles and gas mixing with detailed element yields and metal tracing techniques.

We are developing a series of hydrodynamic simulations to model early galaxy formation in a cosmological context, called the A{\sc eos} project, that focus on tracing individual stars throughout their life cycles and modeling their chemical evolution in unprecedented detail. In this paper, we present our initial simulation. This is a 1 cMpc cubical domain containing a variety of early galaxies. In subsequent simulations, we will perform zoom-ins of small early galaxies that can be regarded as early analogs of the surviving ultra-faint dwarf galaxies. We use the methodology of \cite{Emerick2019} (henceforth called \firstpaper) with updated models and stellar element yields described in Section \ref{sec:methods}. These simulations aim to provide insights into:
\begin{itemize}
    \item metal mixing within the interstellar medium at high redshift,
    \item the origins of the spreads in the observed stellar chemical abundances,
    \item how physically motivated prescriptions for inhomogeneous mixing and abundance scatter can be developed, 
    \item the identification of signatures associated with stars formed from Population III enrichment,
    \item the role of $\alpha$- and $s$-process elements in constraining star formation timescales and stellar ages,
    \item the impact of adopting a star-by-star approach to star formation in simulations.

\end{itemize}
These simulations aim to shed light on the intricate interplay between galaxy formation processes, early nucleosynthetic events, and the stellar chemical abundances we can observe today.

This paper introduces the simulation and the novel handling of star-by-star chemical enrichment. In Section \ref{sec:methods}, we describe the methods implemented in the simulations, focusing on the new or updated physics models. We present an overview of the fiducial simulation in Section \ref{sec:overview} and metal evolution analysis in Section \ref{sec:metalevol}. We conclude in Section \ref{sec:conc}.

\section{Methods} 
\label{sec:methods}

Our simulations use the adaptive mesh refinement (AMR) technique implemented in the code {\sc Enzo}, a grid-based hybrid code that solves the equations of hydrodynamics and gravity \citep{Enzo2014,Enzo2019}. We achieve a physical resolution of 1 pc in regions of high gas density, allowing us to resolve the small-scale structure of the interstellar medium (ISM). The computational grid solves for hydrodynamic quantities, gravitational potential from gas, stars, and dark matter, and evolves star particles with an adaptive particle-mesh N-body solver (Section \ref{subsec:hydro}). Radiative cooling is modeled with a nine-species chemical network (Section \ref{sec:grackle}), and we implement stellar feedback through mass and energy injection from stellar winds and supernovae (Section \ref{sec:feedback}). We model star formation for Population III and Population II stars independently (Sections \ref{sec:SF:PopIII} and \ref{sec:SF:PopII}, respectively), implement their radiation feedback (Section \ref{sec:stellar radiation}), as well as stellar winds and SNe for each population (Sections \ref{sec:stellar feedback:popIII} and \ref{sec:stellar feedback:popII}, respectively). We capture chemical evolution driven by nucleosynthesis from different yield channels, including Pop III CCSNe, Pop II CCSNe, AGB winds, massive stellar winds, and SNIa (Section \ref{sec:stellar yields}). We describe the initial conditions in Section \ref{sec:initialconditions}. 

These methods expand upon those used in \firstpaper\ with the addition of a Pop III star formation and stellar feedback model, which is an updated version of the one used in \citet{Wise2012a}, and a new carefully-chosen set of stellar yields.

\subsection{Numerics}

\subsubsection{Hydrodynamics} \label{subsec:hydro}

Within {\sc Enzo}, we employ the Piecewise Parabolic Method (PPM) for hydrodynamics \citep{ColellaWoodward1984,Bryan1995} and a two-shock approximate Riemann solver. In cases where higher-order methods result in negative densities or energies, the solver progressively falls back to more diffusive Riemann solvers to maintain stability. The total gravitational potential is calculated from gas self-gravity, star particles, and dark matter. We compute self-gravity using a multigrid Poisson solver, while the collisionless star particles are evolved with an adaptive particle-mesh N-body solver at an effective force resolution of approximately 2$\Delta$x, where $\Delta$x is the local cell size.

Mesh refinement is employed to ensure that the thermal Jeans length is resolved by a minimum of 8 cells. Refinement continues in any given region until this criterion is met or the region reaches the maximum resolution of 1 pc. At this maximum resolution, the Jeans length may become under-resolved, potentially leading to artificial numerical fragmentation. To mitigate this, we follow the guideline from \citet{Truelove1997}, which requires the Jeans length to be resolved by at least 4 cells to suppress such fragmentation.


Unlike in \firstpaper, we do not use an artificial pressure floor. Instead, our simulations operate at sufficiently high resolution, allowing us to accurately resolve dense gas regions. This high resolution enables the efficient conversion of dense gas into stars and ensures that stellar feedback processes, such as supernova explosions and stellar winds, effectively disperse the remaining gas.

\subsubsection{Radiative Cooling and Chemistry}
\label{sec:grackle}

We use a slightly modified version of {\sc grackle} \citep{GrackleMethod} to follow both the nine species non-equilibrium chemistry network including H, H$^+$, He, He$^+$, He$^{++}$, e$^{-}$, H$^-$, H$_{2}$, and H$_2^+$ and the effects of radiative cooling and heating from metal lines and an ultraviolet (UV) background. We include the effects of H$_2$ formation on dust using a broken-power law dust-to-gas ratio from \citet{Remy-Ruyer2014} and a \citet{HM2012} UV metagalactic background accounting for self-shielding in the \HI~band and propagating its impact self-consistently to
other photoionization and photodissociation reaction rates and the metal line cooling (see \firstpaper\ for more details). We include a photoelectric heating model for far ultraviolet (FUV) radiation from both individual stars using the same dust-to-gas ratio scaling mentioned above, and a local attenuation approximation.

New in the A{\sc eos} simulations is: 1) the additional contribution of the UV background to the FUV band, 2) the use in the photoelectric heating rate of a constant efficiency parameter instead of one that depends on local gas density, 3) the effects that both FUV and infrared (IR) radiation have on H$_2$ and H${^-}$ reaction rates, and 4) a UV background that has been extended to high redshift in all bands by adopting rates assuming a blackbody radiation spectrum at 3$\times 10^4$~K that turns on at $z = 50$ and is scaled to be continuous with \citet{HM2012} at $z = 10$. The UV background is included to account for the radiation from stars beyond the edges of the simulated volume. In addition, we use an updated Lyman-Werner (LW) background model from both \firstpaper\ and \citet{WiseAbel2012}, adopting the rates at high redshift from \cite{Qin2020}. 

For the photoelectric heating rate, we estimate the efficiency as $\epsilon = 0.05$. Calculating the photoelectric heating efficiency requires a determination of the electron number density $n_e$, which is challenging in dense and neutral regions. In these environments, $n_e$ is largely influenced by the ionization of carbon, dust grains, and polycyclic aromatic hydrocarbons. Our existing chemical network only accounts for electrons contributed by hydrogen, helium, and molecular hydrogen. To address this gap, we considered a power-law relationship for the heating efficiency $\epsilon$ based on the hydrogen number density $n_H$, as derived from the \citet{Wolfire2003} model of $\Gamma_\text{Pe}$ in the solar neighborhood, as also done in \firstpaper: $\epsilon = 0.0148 \times n_H^{0.235}$. For densities ranging from 1 to 10000 cm$^{-3}$, this ranges from $\epsilon$ = 0.015 to 0.12, with typical values around 0.05. We thus take 0.05 as our efficiency.

\subsubsection{Feedback Injection}
\label{sec:feedback}

Stellar winds and supernovae (SNe) are the two sources of both mass and energy feedback included in these simulations. We consider two distinct types of stellar winds: asymptotic giant branch (AGB) winds and winds from massive stars, as well as both core-collapse SNe and Type Ia SNe. The exact mass, metal abundances, and energy deposited by each of these is discussed in Section \ref{sec:feedback}. Here we discuss how each is deposited onto the computational grid.

We have sufficiently high resolution (1 pc) in these simulations to reliably resolve the Sedov-Taylor phase of a majority of our SNe for the typical gas densities in which they explode (see \firstpaper \ and also \citealt{Smith2018b, Hu2019}). For this reason, we include only the thermal energy deposition from these events. Each particle deposits mass and energy feedback over a 2~pc radius spherical region centered on each particle. We use a Monte-Carlo volume overlap calculation to compute the fractional deposition of mass and energy to grid cells that sit on the boundary of the spherical region.

Similarly, stellar winds are deposited in the same 2~pc radius region. The mass loss rates for both AGB and massive star winds are adopted from stellar evolution models,
but are assumed to have fixed velocities and constant loss rates over their AGB phase or lifetime. 

The computational expense of fully resolving fast (10$^3$ km~s$^{-1}$), hot ($10^6$~K) stellar winds that are continually injected onto the grid is too onerous for long-timescale, galaxy-scale simulations, as discussed in \firstpaper. For that reason, we fix the wind velocity for all massive stars to a maximum value of 100 km s$^{-1}$, and fully thermalize the kinetic energy before injection. While this model would reduce the dynamical impact of winds on the evolution of our galaxies, we expect that stellar winds are subdominant to both stellar radiation feedback and SNe, particularly at low metallicity, so we argue this is a reasonable approximation.

\subsection{Star Formation}
\label{sec:SF}

Stars in our simulation form stochastically in cold, dense gas that exhibits a converging flow $\nabla \cdot v < 0$, assuming the local star formation rate is proportional to an efficiency per free-fall time $e_{\rm ff} = 2\%$. In this work, we allow star formation below $T_{\rm thresh} = 500$~K and adopt a high density threshold $n_{\rm thresh} = 10^4$~cm$^{-3}$. Stars are formed when at least 100 M$_\odot$ of gas is available that meets star formation conditions. The initial mass function (IMF) is sampled, stochastically forming stars until the gas reservoir is depleted. We distinguish between Pop III and Pop II star formation based on total gas metallicity. Gas enriched with $Z > 10^{-5} Z_{\odot}$ forms Pop II stars \citep{Ji2014, Chiaki2015,Schneider2012}, while gas below this threshold forms Pop III stars \citep{Tumlinson2006}. 
For Pop III star formation, we place an additional constraint that the molecular hydrogen fraction $f_{\rm H_2} > 0.005$, which is consistent with the $f_{\rm H_2}$ derived from high-resolution simulations of Pop III star formation at our adopted threshold number density $n_{\rm thresh}$ \citep{Susa2014, Kulkarni2021}. We describe the behavior for each population in more detail below.

Our cosmological simulations are \emph{star-by-star} for both Pop III and Pop II stars. This means that in each star formation event, stellar masses are sampled from an adopted IMF and assigned to individual, distinct stellar particles. The exception to this is Pop II stars below 2 M$_\odot$; these stars are aggregated into a single particle during a star formation event because they do not have significant feedback or enrichment on the timescales of these simulations.  

\subsubsection{Pop III Star Formation}
\label{sec:SF:PopIII}

The IMF for Pop III stars remains uncertain \citep[e.g.][]{Bromm13,Latif2022,Klessen2023}. We adopt the same IMF used in \citet{Wise2012a}, which behaves as a \citet{Salpeter1955} IMF with power-law slope $\alpha=-1.3$ above a characteristic mass $M_{\rm char}$ and has an exponential cut-off below $M_{\rm char}$. Motivated by \citet{Hirano17, Bromm13,Yoshida06} and by results from our own trials with varying parameter choices, we adopt $M_{\rm char} = 10$~M$_{\odot}$, with a range of Pop III stellar masses of 1--100~M$_{\odot}$. All Pop III star particles represent individual stars over this mass range. We also run an additional simulation with $M_{\rm char} = 20$~M$_{\odot}$ that we discuss in a separate paper (Brauer et al., in prep). Pop III stars are assigned lifetimes from \citet{Schaerer2002}.

\subsubsection{Pop II Star Formation}
\label{sec:SF:PopII}

Above metallicity $Z > 10^{-5}$~$Z_\odot$, gas is considered to be metal-rich enough to form Pop II stars as sampled from a \citet{Kroupa2001} IMF with a mass range of 0.08--120~M$_{\odot}$. Due to computational constraints, we restrict which stars over this mass range are followed individually to those with $M > 2$~M$_{\odot}$. All stars below this threshold in a star formation event are aggregated together into a single particle. These stars can be combined because they do not have significant feedback or metal enrichment on the timescale of these simulations. Since the low-mass stars with $M \lesssim 1$~M$_{\odot}$ are the only stars that will live to the present day, though, they are key tracers of stellar abundances in present-day low mass dwarf galaxies. We use the zero-age main sequence properties from the PARSEC \citep{Bressan2012,Tang2014} stellar evolution data set to assign stellar radii, effective temperature, surface gravity, lifetimes, and the length of the asymptotic giant branch (AGB) phase, when relevant.

\subsection{Stellar Feedback} 
\label{sec:feedback}

We model detailed multi-channel stellar feedback from each of our stars. Our feedback channels include: core-collapse supernovae (CCSN) 
from Pop III stars, CCSN from Pop II stars, Type Ia supernovae (SNIa), AGB winds, and massive star winds from Pop II stars. We also include stellar radiation followed in three optically-thin bands (IR, FUV, and LW) and \HI, He~{\sc i}, and He~{\sc ii} ionizing radiation followed with an adaptive ray-tracing radiative transfer method including radiation pressure on \HI. These methods are discussed in greater detail below. The yields for each of these events are given in Section \ref{sec:stellar yields}.

\subsubsection{Stellar Radiation}
\label{sec:stellar radiation}

In addition to the UV background, we follow the star-by-star radiation in six bands, separated by photon energy $E_{\rm ph}$. Due to computational constraints, we limit the number of radiation sources---while capturing the vast majority of the photon energy budget of our stars---by restricting radiation to massive stars with $M_* >$8~M$_{\odot}$. 

We follow the \HI\ ($E_{\rm ph} > 13.6$~eV), He\,{\sc i} ($E_{\rm ph} > 24.6$~eV), and He\,{\sc ii} ($E_{\rm ph} > 54.4$~eV) ionizing photons using the \textsc{ENZO+MORAY} adaptive ray-tracing radiative transfer model described in detail in \citet{WiseAbel2011} and \citet{Enzo2014}. Briefly, this method integrates the full equations of radiative transfer, propagating photons mapped onto a \textsc{HEALPix} grid, and adaptively refining once the separation angle between photon packages becomes large. 

In addition, we track the stellar IR ($0.76$~eV~$< E_{\rm ph} < 5.6$~eV), FUV ( $5.6$~eV~$< E_{\rm ph} < 11.2$~eV), and LW ($11.2$~eV~$< E_{\rm ph} < 13.6$~eV) radiation using an optically thin approximation. This allows us to follow local variations in the $H_2$ (LW), $H_2^+$ (IR, FUV, and LW), and $H^-$ (IR) photodissociation rates from each band, in addition to the localized photoelectric heating from stellar FUV radiation.

\subsubsection{Pop III Stellar Feedback}
\label{sec:stellar feedback:popIII}

We use the table of binned photon counts from \citet{HegerWoosley2010} with the lifetimes in \citet{Schaerer2002} to compute the constant photon fluxes for our Pop III stars in each radiation bin (IR, FUV, LW, and \HI, He\,{\sc i}, and He\,{\sc ii} ionizing radiation) as a function of stellar mass. In practice, this is implemented using a piece-wise polynomial fit to these tables. 

Pop III CCSN with 10 M$_{\odot} < M_* < 100$~M$_{\odot}$ explode with a fixed energy of 10$^{51}$~erg.

\subsubsection{Pop II Stellar Feedback}
\label{sec:stellar feedback:popII}

We use the \textsc{PARSEC} \citep{Bressan2012} grid of stellar evolution tracks to set the lifetime of each star and the start time and length of the AGB phase, if present. This is also used to set the stellar effective temperature, surface gravity, and radius---each of which remain fixed at their zero age main sequence values---which are in turn used to set the radiation properties of each star. Photon fluxes in each radiation band are determined using the OSTAR2002 \citep{Lanz2003} grid of O-type stellar models. 

However, this table does not have complete coverage over all possible stellar properties encountered in these simulations, particularly for stars below about 15~M$_{\odot}$ and very massive stars with sub-solar metallicity. For stars off of the grid, we adopt a blackbody spectrum with rates scaled to be continuous with the OSTAR2002 grid (see Appendix B of \firstpaper). Ionizing photon energies are taken to be the average ionizing photon energy for the corresponding blackbody spectrum of each star. Stellar wind velocities are fixed to 20 km s$^{-1}$ for AGB stars ($M_* < 8$~M$_{\odot}$), and 100 km s$^{-1}$ (our wind velocity ceiling, see Section~\ref{sec:feedback}) for massive stars ($M_* > 8$~M$_{\odot}$). CCSN occur for stars between 8--25 M$_{\odot}$ with an energy of 10$^{51}$~erg, and we assume stars more massive than 25~M$_{\odot}$ directly collapse into black holes with no mass or energy feedback \citep{Limongi2018}.

In \firstpaper, we used a power-law to describe the delay time distribution (DTD) for the occurence of SNIa assuming a single formation channel. We update our prescription by adopting the standard DTD from \citet[][their model A1]{Ruiter2011}, which provides the total SNIa DTD as the sum of four different channels: 1) double degenerate scenario, 2) single degenerate scenario, 3) helium-rich donor scenario, and 4) a sub-Chandresekhar mass scenario. 

Our SNIa prescription utilizes the initial mass-to-final mass relation of \citet{Cummings2019} to assign masses to the corresponding white dwarf particles once stars below 8 M$_{\odot}$ have reached the end of their lives. Following \firstpaper, we assume that stars with initial masses of 3--8 M$_{\odot}$ form white dwarfs capable of exploding as SNIa. Given this, the DTD, and our IMF, the fraction of stars capable of forming SNIa progenitors that will explode in a Hubble time is $0.1508$ (see Eq. 2 in \firstpaper). We pre-tabulate the cumulative probability distribution for both the total DTD and each underlying DTD. When a white dwarf forms, we use a random number draw over the total DTD to set the time (if any) that each SNIa candidate will explode and make a separate random number draw to decide which type it will be. For simplicity, we treat the total energy output for each SNIa as being the same 10$^{51}$~erg and differentiate them only by their yields (see Section~\ref{sec:yields:Ia}).

\subsection{Stellar Yields}
\label{sec:stellar yields}

We pay careful attention to capturing the detailed chemical evolution driven by nucleosynthesis from distinct yield channels in both Pop III and Pop II stars, as detailed below. For a figure showing all yields as a function of progenitor mass, see Appendix \ref{sec:appendix_yields}.

In total, we track 10 individual metal abundances (in addition to H, He, and the total metallicity): C, N, O, Na, Mg, Ca, Mn, Fe, Sr, and Ba. This well samples elements from each nucleosynthetic channel, in addition to capturing elements with different mass and metallicity dependence in each channel. 
Oxygen (O), magnesium (Mg), and calcium (Ca) are produced predominately in CCSN and show a noticeable evolution with SN progenitor mass, tracing short-timescale ($10~$Myr) chemical evolution. Iron (Fe) is produced in both core collapse and SNIa, and the relative abundances of O, Mg, and Ca to Fe trace the evolution between these two nucleosynthetic sources on timescales of 0.1 to 1~Gyr. Nitrogen (N), strontium (Sr), and barium (Ba) trace $s$-process enrichment in low-mass AGB stars on timescales of 0.1 to 1~Gyr. N and Ba trace the most massive (4--8~M$_{\odot}$) AGB stars, while Sr traces the less massive ($<4$~M$_{\odot}$) ones. Sodium (Na) is produced in AGB stars as part of the NeNa cycle, mostly in intermediate-mass ($\sim 4$M$_{\odot}$) AGB stars. Carbon (C) is significantly produced in both low-mass AGB stars and CCSN, and is also an important tracer of early Pop III enrichment (in the form of carbon-to-iron ratios). Manganese (Mn) is predominantly formed in SNIa.  These elements are readily observed in stellar spectra, with the exception of N and O, with O being the primary tracer of gas-phase abundances. 

In addition, we follow tracers tracking the total metal mass in each cell from each yield source in our chemical evolution model: Pop III CCSN, 
AGB winds, massive star (M > 8 M$_{\odot}$) winds, Pop II CCSN,  and SNIa. Our SNIa prescription (Sect.~\ref{sec:yields:Ia}) includes four metal tracers for different SNIa progenitor types. We additionally include an $r$-process yield tracer 
to allow post-processing of $r$-process abundances. 
In total---counting the total metallicity tracer---we follow 20 metal tracers.

\subsubsection{Pop III Yields}
\label{sec:popiii yields}

For the CCSNe from Pop III stars (10 M$_{\odot}$ < $M_*$ < 100 M$_{\odot}$), we adopt the yields from \citet{HegerWoosley2010} with standard 0.1 mixing.
While the exact fate of Pop III stars in the range 70--120~M$_{\odot}$ is uncertain---with some possibly exploding as CCSN, 
and others undergoing direct collapse with no yield return---it is reasonable to approximate that all of these stars, at least up to 100~M$_{\odot}$, end their life in a CCSN event \citep{Woosley2017}. 


\subsubsection{Pop II Yields}
\label{sec:popii yields}

For AGB winds ($M_{*} < 8$~M$_{\odot}$), we adopt the yields of \citet{Cristallo2015}, with 8 grid points in $M_* \in [1.3,6.0]$~M$_{\odot}$ and 10 in $Z_* \in [1 \times 10^{-4},0.02]$. For the winds and CCSNe yields of massive stars, we adopt \citet{Limongi2018}, with 9 grid points over $M_* \in [13,120]$~M$_{\odot}$ and four in $Z \in [3.236\times 10^{-5},0.01345]$. Stellar yields are interpolated linearly between mass and metallicity grid points in each of the tables. For stars with masses outside the mass range sampled by the yield tables, we adopt the abundance ratios of the nearest grid point and scale the yield mass linearly with stellar mass. Yields for stars with metallicities outside the covered range are taken to be the same as the yield of the closest grid point with no extrapolation in $Z$. 

The yield models from \citet{Limongi2018} are presented for three different stellar rotations. Rather than accounting for these differences live in our simulations, we adopt a pre-computed mixture model representing a population-averaged yield set using the metallicity-dependent stellar rotation population fractions from \citet{Prantzos2018}. In order to fully sample the variations in \citet{Prantzos2018} with metallicity, we pre-compute an interpolated mixture model using an additional three evenly log-spaced metallicities in between the four existing grid points for a total of 13 metallicities.

We adopted our particular set of elemental yields based on comparing the results of a one-zone galactic chemical evolution model as applied to a Milky Way mass galaxy. While these yields generally produced reasonable agreement in this model as compared to observations in [X/Fe] vs.\ [Fe/H] space, Mg is noticeably under-produced in [Mg/Fe] at all [Fe/H], while other $\alpha$-elements tend to agree well with observed Milky Way chemical abundances. Given the importance of accurate observational comparisons in our simulations, we applied a uniform factor of 2.2 to the Mg yields from all massive stars, which brings the model into agreement with observed Mg abundances. This is discussed in Appendix \ref{sec:appendix_yields}.

\begin{figure*}[hbtp!]
\center
\includegraphics[width=\linewidth]{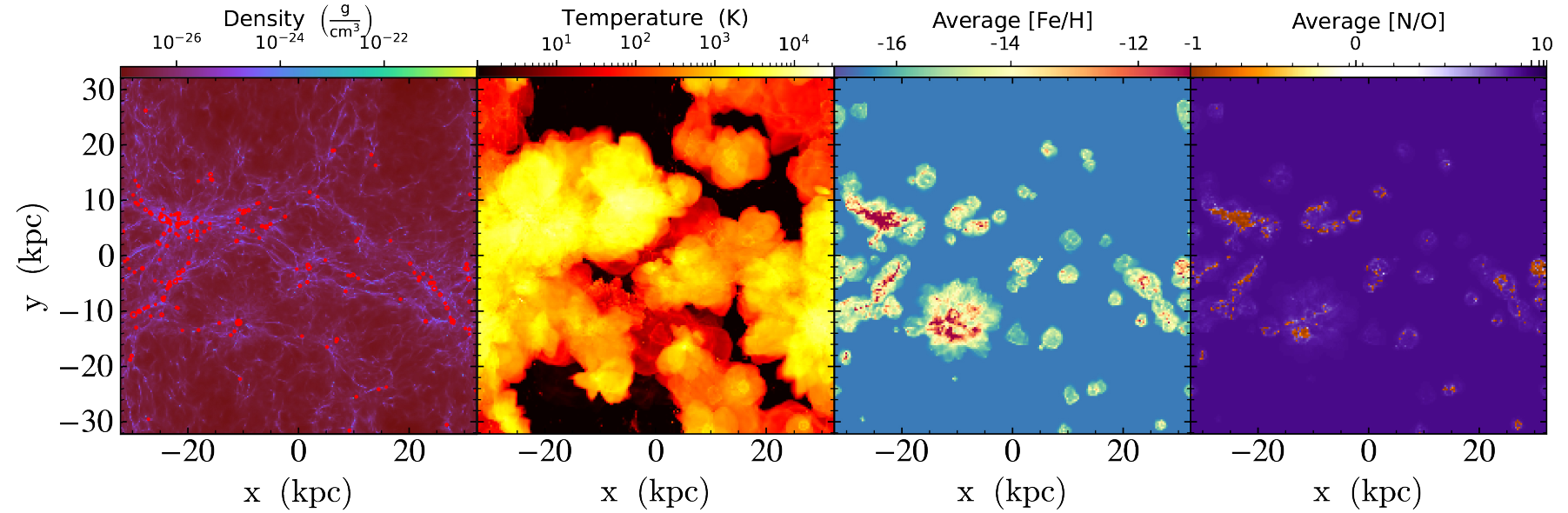}
\caption{Density-weighted averages of the gas density, temperature, [Fe/H], and [N/O] of the fiducial A{\sc eos} simulation at redshift $z=14.5$. At this redshift, 91 halos have begun forming stars. Each star-forming halo is circled in red in the gas density projection. 
\label{fig:fullbox}}
\end{figure*}

\subsubsection{SNIa Yields}
\label{sec:yields:Ia}

As discussed in \ref{sec:stellar feedback:popII}, we use a combined DTD from four different sources of SNIa, each with potentially unique abundance signatures. However, given the uncertainty in yields from each of these sources, we opt instead to make the assumption that there is one yield pattern for each source, which allows us to post-process the abundance patterns from each channel separately. Live in the simulation, we assume a single abundance pattern for \textit{all} SNIa from \citet{Thielemann1986} and track the contribution of each SNIa type to the total metallicity as a separate passive scalar tracer field. Given this, and knowing the total number of each SNIa type that has occurred in the simulation, one can arbitrarily rescale the abundance patterns for each SNIa type. Doing so implicitly assumes that the yields for each SNIa do not affect the dynamical evolution of our galaxies (which, in reality, it may, for example by influencing cooling through the Fe atomic line cooling), since we do not account for local cooling variations due to individual elemental abundances in our simulations.


%

\subsection{Initial Conditions}
\label{sec:initialconditions}

For the simulations presented here and in our companion paper (Brauer et al., in prep), we use the same initial conditions as \citet{Skinner20}. These initial conditions were created with \textsc{music } \citep{Hahn2011} at $z = 130$ and use the cosmological parameters from the Planck collaboration best fit \citep{Planck2014}.

\section{Overview of Fiducial Simulation}
\label{sec:overview}

Our initial simulation has a co-moving (1 Mpc)$^3$ volume, simulated from redshift $z = 130$ to $z=14.5$  ($\sim300$ Myr after the Big Bang). It has a root-grid resolution of $256^3$, a dark matter resolution of 1840 M$_\odot$, and a physical 1 pc resolution of the gas at the finest scales. All Pop II stars with masses greater than 2 M$_\odot$ and all Pop III stars are represented by single star particles. The stars and gas have 20 metal tracer fields tracing 10 individual metal abundances and several yield sources (see Section~\ \ref{sec:stellar yields}). Projections of the fiducial volume can be seen in Figure \ref{fig:fullbox}.

Because computational limitations do not allow us to simulate the full domain beyond redshift $z \sim14$, our initial analysis focuses on the first 300 Myr of the Universe. We also run an additional A{\sc eos} simulation with a different Pop III IMF and several comparison simulations without individual stellar feedback; these simulations are discussed and compared in our companion paper (Brauer et al., in prep.). Future work will include zoom-in simulations of ultra-faint dwarfs that will run until reionization.

\begin{figure}[b]
\center
\includegraphics[width=0.95\linewidth]{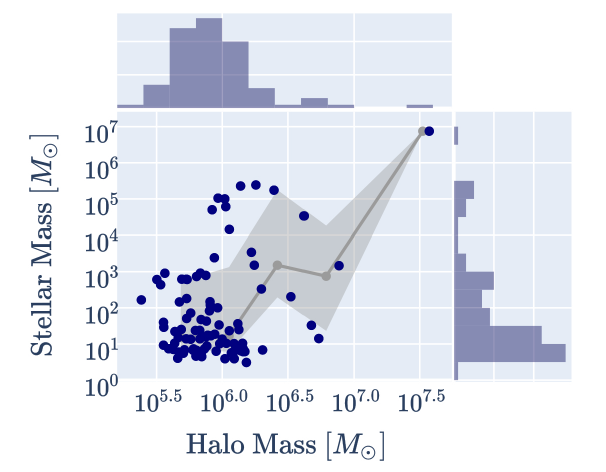}
\caption{The individual stellar masses and halo masses of every star-forming halo in the simulation at redshift $z=14.5$ are shown in navy. The mean stellar-mass to halo-mass relation for these galaxies with 16th-84th percentile scatter is shown in grey.
\label{fig:MsMh}}
\end{figure}

At redshift $z = 14.5$, the full 1 Mpc volume contains 91 star-forming halos of at least $M_{\text{halo}} > 2\times10^5$ M$_\odot$ (our resolution limit) with a total of about 250,000 star and stellar remnant particles (see Figure \ref{fig:fullbox}). Pop III star formation begins at $z \sim 28$ (110 Myr after Big Bang) and Pop II star formation begins a bit after $z \sim 22$ (160 Myr after Big Bang). 
The mass of Pop II stars, defined as stars with total metals $Z \geq 10^{-5}$~$Z_\odot$, overtakes the mass of Pop III stars around $z \sim 17$. Figure \ref{fig:MsMh} shows the stellar masses and halo masses of every star-forming halo in the simulations at $z=14.5$ and the stellar-mass to halo-mass relation.

Some galaxies in our simulations exhibit unexpectedly high stellar-to-halo mass ratios (such as $M_* = 10^{5}\,M_\odot$ for $M_{\text{halo}} = 10^{6}\,M_\odot$). Abundance matching relations are very poorly calibrated for low-mass halos, but they predict closer to $M_* = 10^{2} - 10^{3}\,M_\odot$ or even lower for a $10^{6}\,M_\odot$ halo \citep[e.g.,][]{Behroozi13}. Most of our galaxies are indeed low in stellar mass, and we have significant scatter in stellar mass, which is expected with lower halo masses \citep{GarrisonKimmel17}. For the massive outliers, their high stellar-to-halo mass ratios may be due in part to environmental factors. Two of the outlier galaxies are the externally enriched galaxies near the central galaxy (see Section \ref{sec:extenrich}), and their total halo mass is affected by their orbit around a more massive galaxy, allowing for dark matter stripping (a lesser version of the stripping processes described by \citet{Moreno22}).

The unexpectedly high stellar-to-halo mass ratios may also be due to our star-by-star IMF sampling approach. Recently, \citet{Jeon2024} found that star-by-star sampling leads to higher stellar masses in low-mass halos (e.g., $10^8\,M_\odot$) compared to traditional single stellar population (SSP) particles, which inject feedback more burstily and suppress star formation more effectively. \citet{Jeon2024} demonstrated that the SSP method results in burstier, stronger feedback, while our star-by-star method, with its weaker feedback, allows for higher stellar masses. They also argue that star-by-star sampling more accurately matches observations and accounts for the high scatter in the $M_{\text{halo}}$-$M_*$ relation at low halo masses. Similarly, \citet{Andersson24} reports a systematic increase in stellar mass for a given halo mass compared to classical galaxy models (see Figure 1 of \citet{Andersson24}). In our other recent \Aeos paper \citep{Mead2024b}, we observe that a single SN disrupts star formation in most galaxies, but with this effect diminishes around $10^7\,M_\odot$ where our outliers lie.

\begin{figure}[t]
\center
\includegraphics[width=0.95\linewidth]{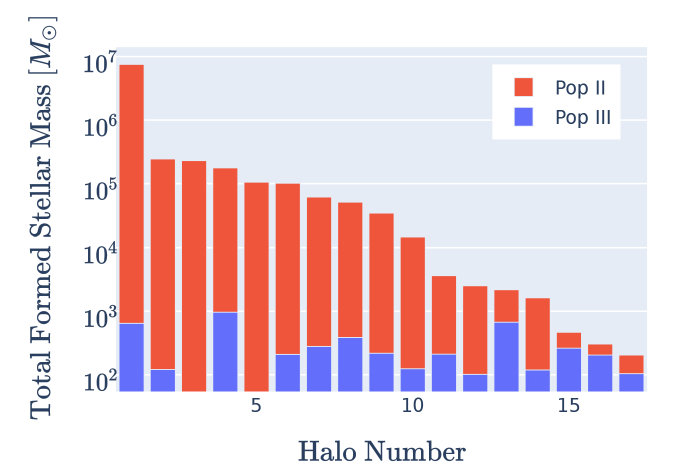}
\caption{For every star-forming halo with Pop II stars, we show the distribution of Pop II vs.\ Pop III stellar mass throughout the history of the simulation (until redshift $z =14.5$). Two halos, Halo 3 and Halo 5, were externally enriched in metals by Halo 1, so their star formation began with Pop II.
\label{fig:popIIgal}}
\end{figure}

Of the 91 star-forming halos, the vast majority are tiny. 60 of the halos contain stellar masses of $M_{*} \leq 100$~M$_\odot$, while only one galaxy has $M_{*}>10^6$ M$_\odot$. 17 of the galaxies have begun forming Pop II stars by $z\sim14.5$. For each of these galaxies, the distribution of Pop II vs.\ Pop III stellar mass is shown in Figure \ref{fig:popIIgal}. Galaxies typically undergo multiple episodes of Pop III star formation before transitioning to Pop II stars. The extent of Pop III star formation required for this transition is strongly influenced by the halo mass, which affects the galaxy’s ability to retain metals, as well as the star formation activity of neighboring galaxies. For example, in Halo 1 of Figure \ref{fig:popIIgal}, approximately 1,000 solar masses of Pop III stars formed over an 80 Myr period before the onset of Pop II star formation. Other galaxies exhibited a different amounts of Pop III star formation, ranging from none or only a few hundred solar masses -- particularly in cases of external enrichment, more massive halos, or proximity to other star-forming galaxies -- to up to about 1,000 solar masses in galaxies that began star formation earlier than their surroundings or were relatively isolated. This variation highlights the significant role that both intrinsic halo properties and environmental factors play in determining the transition from Pop III to Pop II star formation.

\begin{figure*}[tbp!]
\center
\includegraphics[width=0.7\linewidth]{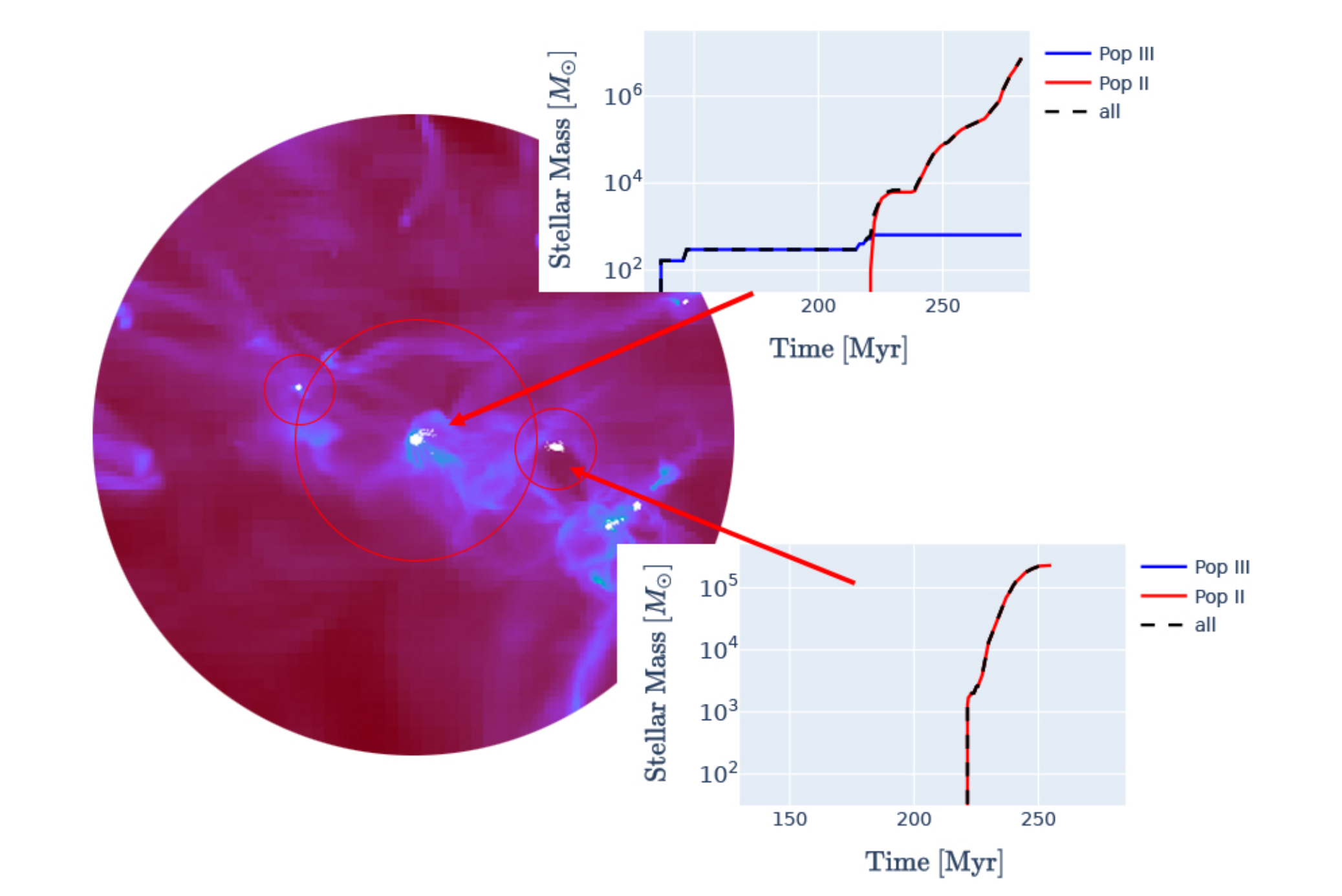}
\caption{Two galaxies around Halo 1 (Halos 3 and 5 from Figure \ref{fig:popIIgal}) were externally enriched in metals by outflows from Halo 1 sufficiently early that their star formation started with Pop II. Here we show the cumulative star formation history of Halo 1 and Halo 3. The virial radius of Halo 5 is also shown to the left. The physical extent of the gas projection is about 4 kpc.
\label{fig:extenrichment}}
\end{figure*}

\subsection{External Enrichment and Systems of Galaxies} \label{sec:extenrich}

Two galaxies, Halo 3 and Halo 5, were externally enriched in metals by feedback-driven outflows from a larger galaxy, Halo 1. This caused their star formation to begin with Pop II stars. This is seen in Figure \ref{fig:popIIgal} and shown in more detail in Figure \ref{fig:extenrichment}. Both Halo 3 and Halo 5 sit just outside the virial radius of Halo 1 and will merge in the future, but even before this future merger they all share gas and metals.

Due to the shallow potentials of their small halos, the A{\sc eos} galaxies struggle to retain their gas and metals \citep{Mead2024b}. Every SN explosion, in the smallest halos, blows out the galaxy's gas, causing neighboring galaxies to freely share metals and gas between themselves. This leads to behavior like the external enrichment shown in Figure \ref{fig:extenrichment}. 

In these simulations, the halos of early galaxies ($z \gtrsim 14$) are all low-mass (generally $M_{\rm dm} \lesssim 10^7$ M$_\odot$). Any individual galaxy experiencing a SN explosion thus strongly interacts with any galaxies around it (within $\sim 5$ physical kpc). This raises the question of how to treat early galaxies like Halos 1, 3, and 5: as individuals or as systems. Based on the communal sharing of gas and metals between the galaxies in these simulations, we suggest that the earliest galaxies be thought of not as individuals but as systems that contribute their metals to a common reservoir and evolve together.

Examples of external enrichment to the degree of a galaxy beginning its star formation with Pop II stars is still rare, however, as evidenced by Figure \ref{fig:popIIgal}. In this simulation, there are two examples of galaxies that begin star formation with Pop II out of 17 galaxies that have begun forming Pop II stars. This depends strongly on the amount of clustering. The two externally enriched galaxies exist within a few physical kpc of the largest galaxy in the simulation, and will likely merge with the central galaxy in time. So while clusters of small galaxies within $\sim 5$ physical kpc may be considered systems rather than individuals, this does not hold for more isolated early halos. 

We also look at how the metallicities differ for stars formed via external enrichment. When looking at first-generation Pop II stars in the galaxies shown in Figure \ref{fig:extenrichment}, the metallicities differ between the externally enriched halos and the central halo, but can be both higher metallicity or lower metallicity. The first Pop II stars in Halo 1 form at about [Fe/H] = $-5$, while the first stars in Halo 3 form at about [Fe/H] = $-4.5$. When SNe blow out the metals from Halo 1, it actually results in Halo 3 having a higher metal density. On the other hand, Halo 5 is about twice as far away ($\sim 3$ kpc), and when it starts forming stars later at 260 Myr its stars have [Fe/H] = $-5.3$. In this case, the metals became diffused during the transfer.

\section{Individual Stellar Chemical Abundances}
\label{sec:metalevol} 

We track 10 individual metal abundances, as described in Section \ref{sec:stellar yields}.
These elements trace different nucleosynthetic channels with stars of different mass and metallicity within any given channel. The most important novelty of the A{\sc eos} simulations is the ability to trace detailed metal enrichment from individual stars in small galaxies -- we do this in a cosmological context with radiative transfer and include enrichment from Pop III stars. This star-by-star resolution is necessary to uncover scatter and structure in stellar chemical abundance space.

\subsection{Structure in Chemical Abundance Space Corresponding to Galactic Origins of Stars} \label{subsec:abundorigins}

\begin{figure*}
\center
\includegraphics[width=\linewidth]{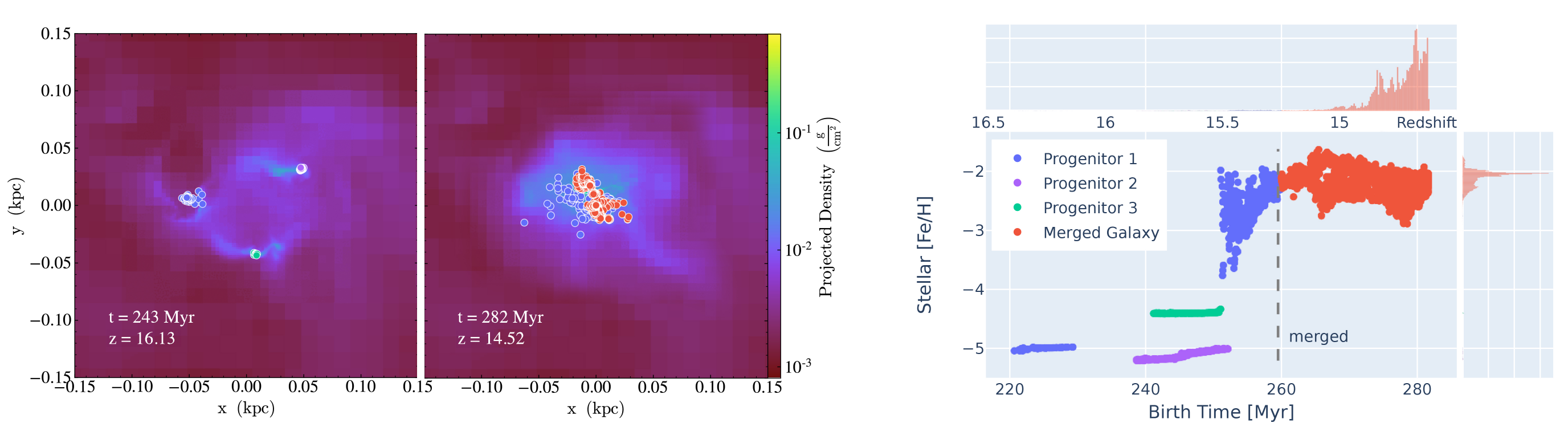}
\caption{{\em (Left)} Around 160 Myr after the Big Bang, three small galaxies merge in the simulation. Each progenitor enters with its own stars, with colors labeled in the right panel, and the merged galaxy continues to form stars. ({\em Right)} We show the birth time and metallicity of each star in this system; the galaxies formed stars at slightly different metallicities prior to merging. Each star is colored by its galaxy of origin. We investigate more chemical abundances in Figure \ref{fig:abund} to show that structure in chemical abundance space can correspond to the origins of the stars. 
\label{fig:originex}}
\end{figure*}

As a proof of concept, consider the merger of three galaxies shown in Figure \ref{fig:originex}. In simulation, three early galaxies merge at $z \sim 15$. At the time of merger, these galaxies contain only 6000 M$_\odot$, 2000 M$_\odot$, and 15000 M$_\odot$ of stellar mass respectively before merging, and after merging experience bursty star formation until the merged halo has a stellar mass of $10^7$ M$_\odot$ at the end of the simulation (Halo 1 at $z=14.5$). This is a clustered region of the simulation.

\begin{figure*}
\center
\includegraphics[width=0.8\linewidth]{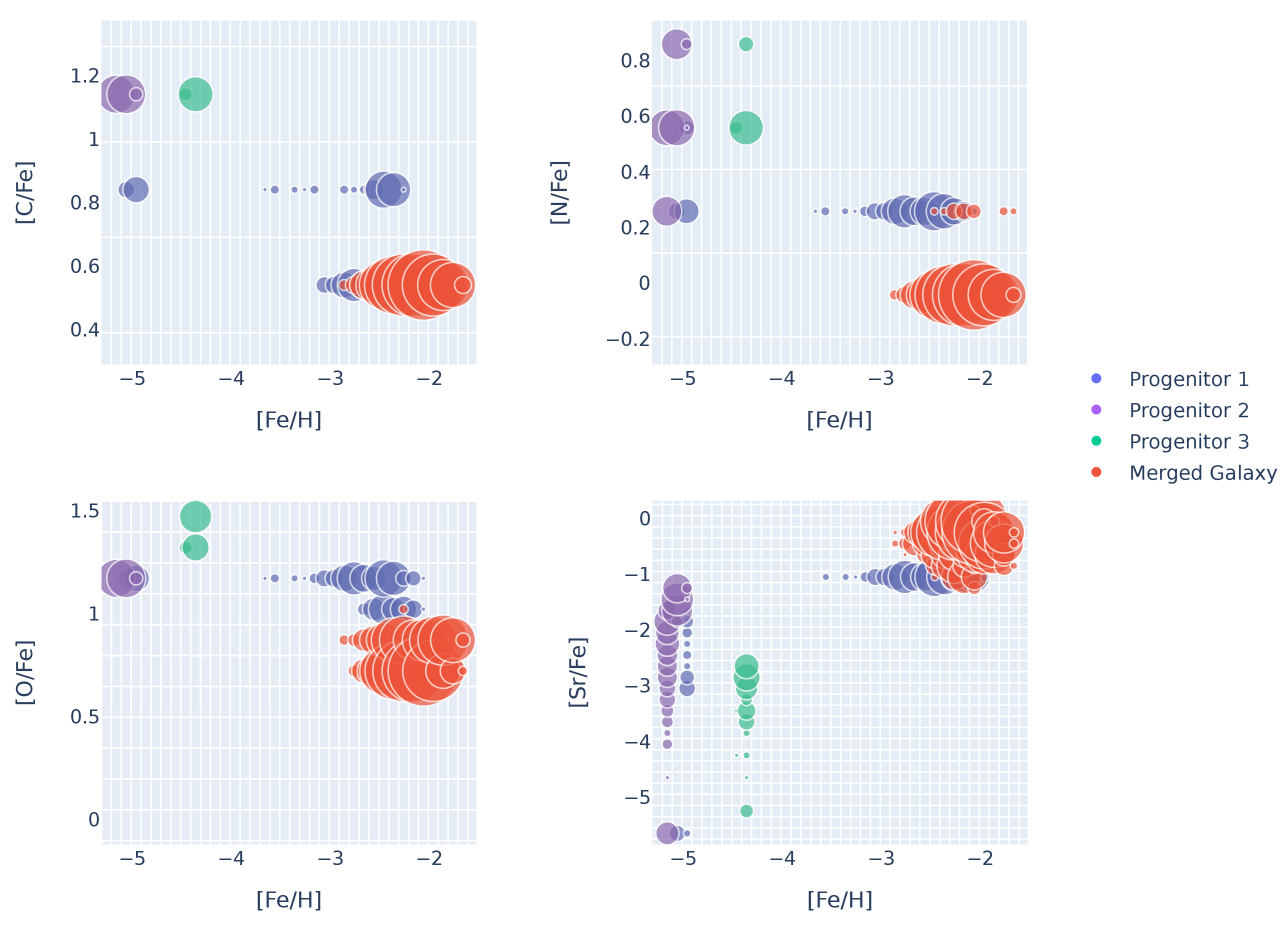}
\caption{Example of structure in chemical abundance space. For the system of merging galaxies shown in Figure \ref{fig:originex}, we plot the chemical abundances of stars (only showing low-mass stars that will still be alive at present-day), colored by whether the star formed in Progenitor 1 {\em (blue)}, Progenitor 2 {\em (purple)}, Progenitor 3 {\em (green)}, or later in the fully merged galaxy {\em (orange)}. The data is binned on expected observational uncertainty for stars in the Milky Way halo today (the grey-white grid). The size of the dot in each bin is proportional to the log of the number of stars in that bin. In this simple example, the location of stars in multi-dimensional chemical abundance space clearly corresponds to which halo the stars were born in.
\label{fig:abund}}
\end{figure*}

Figure \ref{fig:originex} shows the individual star particles of these galaxies, colored according to their progenitor galaxy. As the galaxies merge, star formation increases and the merged galaxy continues to form stars (shown in orange). Here, ``merged'' means that the star-forming regions are indistinguishable from one another at the physical scale of 0.1 kpc, but note that the regions are still oscillating and the halos could still move out and back into Halo 1 before becoming permanently indistinguishable. The projections and stellar birth times shown in Figure \ref{fig:originex} can only ever be known in a simulation, never in any observed data.

The stellar chemical abundances shown in Figure \ref{fig:abund}, however, \textit{could} be measured in observed stars with spectroscopy. This figure shows the [Fe/H], [C/Fe], [N/Fe], [O/Fe], and [Sr/Fe] chemical abundances of the low-mass Pop II stars that would survive to the present day. To mimic the inherent limitations of observations, the abundances are binned according to typical observational abundance uncertainties for stars in the Milky Way halo -- here, we use the typical uncertainties of $\sim$0.1 dex for Fe, $\sim$0.2 dex for Sr, $\sim$0.1-0.2 dex for O (higher at more metal-poor), and $\sim$0.3 dex for N and C. These uncertainties are based on uncertainties of Milky Way stars from JINAbase \citep{JINAbase}. The size of the dot in each bin is proportional to the log number of stars in that bin. Once again, we color the stars according to their progenitor galaxy.

At this point in the simulation in this system, the most important nucleosynthetic sources for the elements in these stars are CCSN (both Pop II and Pop III) with a minimal contribution from AGB winds. In the last $\sim 15$~Myr of the simulation, Pop II CCSN take over as the dominant source of metals in these galaxies. For the elements shown in Figure \ref{fig:abund}, carbon has the greatest AGB contribution, but the mass of carbon from AGB stars is still almost four orders of magnitude lower than the mass from CCSN (see Section \ref{subsec:CEMP}). The spread in strontium relative to iron is due to strontium's more significant production in Pop II AGB winds and CCSN compared to Pop III CCSN (see Figure \ref{fig:yields}), exacerbated by the overall extremely low amounts of strontium in the gas ($\sim 10^{-4}$~M$_\odot$ of strontium in the entire system). Note that strontium is primarily a tracer of AGB winds but forms in small amounts in CCSN due to weak s-process, and the metal content of Pop II stars provide seed nuclei that Pop III stars lack. As time continues to pass, winds and SNIa will become significant and contribute to more scatter, even within individual progenitor galaxies (such as the spread in strontium seen in Figure \ref{fig:abund}). There is also a spread seen in [N/Fe] in the different progenitor galaxies, particularly Progenitor 2. This is due to yield differences from different-mass CCSNe, producing an intrinsic source of scatter.

Figure \ref{fig:abund} demonstrates that structure exists in chemical abundance space and can correspond to the progenitor origins of the stars. In this simple example, the stars from different progenitor galaxies visibly exist in different regions of chemical abundance space. The multi-dimensional abundance space contains rich information about the environments in which the stars formed. In this way, stellar chemical abundances encode information about hierarchical galaxy formation.

In this example, the abundance scatter corresponds to differences in early galaxy formation, but scatter can also be due to differences in nucleosynthetic yields (as we have already alluded to when discussing the spreads of strontium). The current simulations have only run to $z=14.5$, but the next iteration of the A{\sc eos} simulations will run to reionization. This will capture a rich history of different nucleosynthetic events including CCSN, SNIa, $s$-process elements from AGB stars, and $r$-process events that will be included in post-processing. With the level of detail shown in Figures \ref{fig:originex} and \ref{fig:abund}, we will be able to identify how the spreads of abundance distributions differ with variations in galaxy evolution (e.g., merger history, star formation history), and also with variations in the different nucleosynthetic events experienced by each galaxy (e.g., different amounts of barium produced in $s$- vs.\ $r$-process events or stochastic differences in yields from single sources).

\subsection{CEMP Signature in First Generation Pop II Stars} \label{subsec:CEMP}

Carbon-enhanced metal-poor (CEMP) stars are a unique and crucial population in the study of stellar evolution and galactic archaeology. Observationally, these stars are characterized by their unusually high carbon-to-iron ratios ([C/Fe] > +1.0) in comparison to typical metal-poor stars. CEMP stars are of particular interest because they offer significant insights into the early universe, the processes of nucleosynthesis, and the formation of the first stars and galaxies \citep{Beers2005}.

In this system of galaxies (see Figures \ref{fig:originex} and \ref{fig:abund}), first generation Pop II CEMP stars are observed exclusively at extremely low metallicities. The highest stellar metallicity for these CEMP stars is [Fe/H] = -4.3, while the median metallicity of CEMP stars in this halo is [Fe/H] = -5.

More broadly in the simulation, about half of low-metallicity stars that would survive to present-day exhibit a CEMP signature. Below [Fe/H] = -4, 54\% of the low-mass stars (sub-solar mass stars which will survive to present-day) in the simulation present a CEMP signature. The fraction of stars with a CEMP signature decreases to 29\% of the low-mass stars below [Fe/H] = -2.5 and 2\% of the low-mass stars below [Fe/H] = -1.5. This is generally in agreement with observations of metal-poor stars that suggest a significant fraction of metal-poor stars are carbon-rich \citep{Norris2013}, a fraction that increases with decreasing metallicity and may be as high as $\sim$80\% for stars below [Fe/H] = -4 \citep{Placco2014}. 

\begin{figure}[tb]
\center
\includegraphics[width=0.95\linewidth]{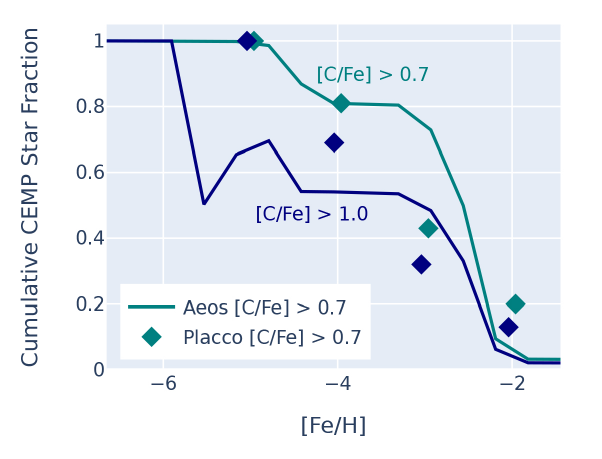}
\caption{The frequency of CEMP stars (with [C/Fe] > 0.7 or [C/Fe] > 1.0) as a function of metallicity (shown in green and blue, respectively). The low-mass A{\sc eos} stars are shown as a line, and observations from \citet{Placco2014} are shown as diamonds.
\label{fig:CEMP}}
\end{figure}

The plot in Figure \ref{fig:CEMP} provides a detailed view of the CEMP fraction across a range of metallicities for all the low-mass stars in the simulation. It shows the fraction of stars with [C/Fe] > 1.0 (blue) and [C/Fe] > 0.7 (green) as a function of [Fe/H]. At the lowest metallicities ([Fe/H] < -5), nearly all low-mass stars in the simulation exhibit a CEMP signature, in alignment with the idea that carbon-enhancement is common in metal-poor environments due to enrichment from the first stars (Pop III) through faint supernovae or winds. As metallicity increases beyond [Fe/H] $\sim$ -4, the fraction of CEMP stars drops significantly, reflecting the shift from environments dominated by Pop III chemical signatures to those where normal Pop II star formation has begun occurring.

In Figure \ref{fig:CEMP}, we also compare to the observed frequencies of CEMP stars in the Milky Way \citep{Placco2014}. At the lowest metallicities, we under-predict CEMP stars with [C/Fe] > 1.0 when compared to Milky Way observations. This could indicate the need for a Pop III IMF that favors higher-mass Pop III, because higher mass CCSNe produce ejecta with a higher ratio of carbon to iron \citep[see Figure \ref{fig:yieldsFe} and ][]{HegerWoosley2010}. Our Pop III IMF has a characteristic mass of $M_\text{char} = 10$ M$_\odot$, which is on the lower end of expectations for Pop III IMF. It could also indicate too-efficient mixing in the simulation volume, which could dilute the carbon enhancement in localized star-forming regions. We note, however, that the low-metallicity end of observations is uncertain due to limited data and the simulated abundances generally agree with observed rates of CEMP stars with [C/Fe] > 0.7.

\subsection{Metallicity Floors from Pop III Enrichment}

The A{\sc eos} simulation differs from other star-by-star chemical enrichment simulations because it includes a Pop III enrichment model in a cosmological context. Simulations without a Pop III model, such as EDGE \citep{Andersson24,Agertz2020}, assume a metallicity floor resulting from Pop III enrichment rather than explicitly modeling Pop III feedback and enrichment. This metallicity floor is generally Z = 10$^{-3}$ or 10$^{-4}$ Z$_\odot$ \citep{Wise2012a,Jaacks18}; in the case of EDGE, they add Z = 10$^{-3}$ Z$_\odot$ to the oxygen field, corresponding to an oxygen metallicity floor of about [O/H] = -2.7 \citep{Agertz2020}.

\begin{figure}[t]
\center
\includegraphics[width=0.95\linewidth]{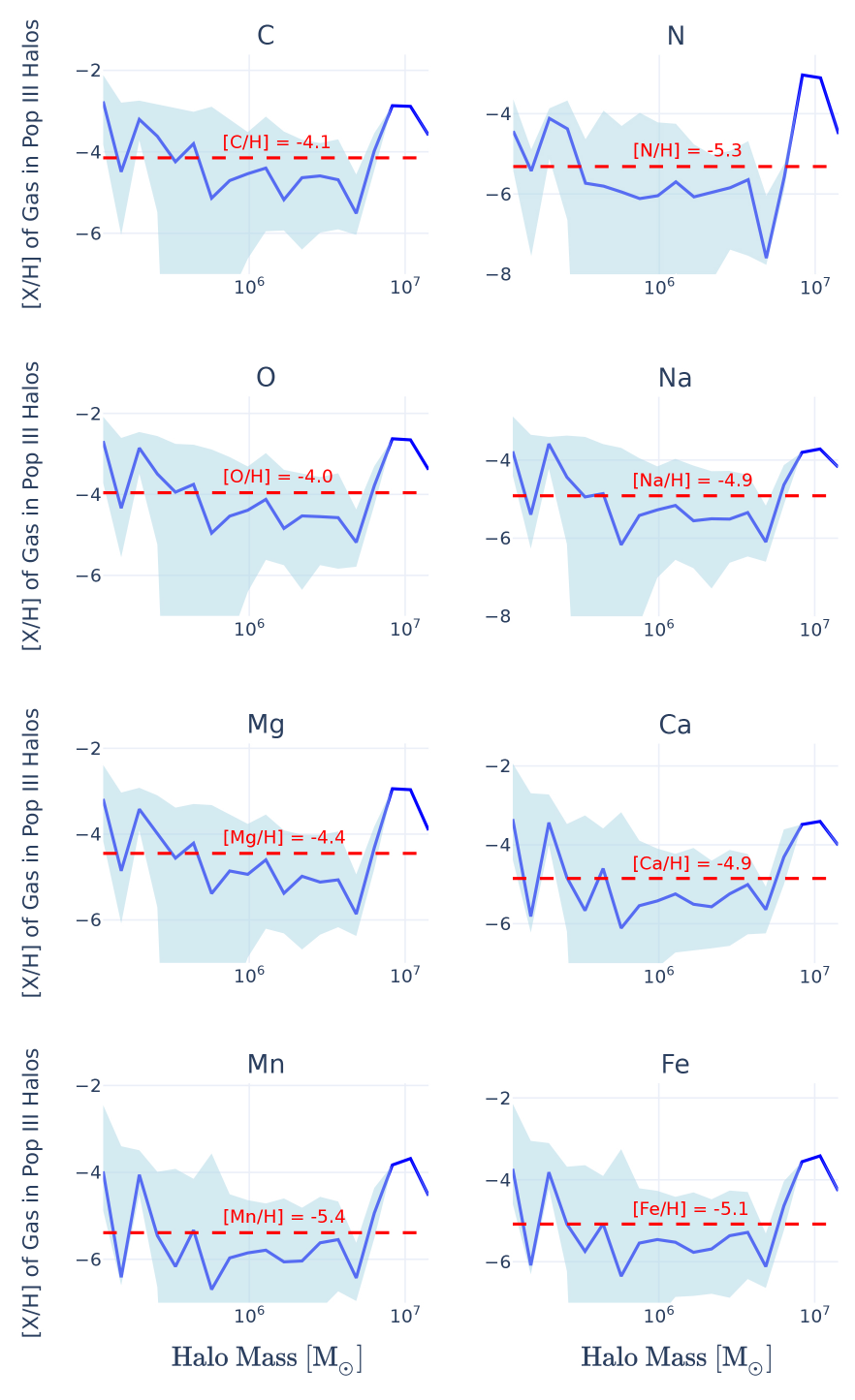}
\caption{The [X/H] of the gas in galaxies that have formed only Pop III. This plot summarizes 270 galaxies from five different snapshots, equally temporarily spaced from about 150 Myr to about 300 Myr. The median and the 16th to 84th percentile scatter at each halo mass are shown in blue. Each metal is fit with a flat line, shown in red, to estimate a metallicity floor from Pop III enrichment.
\label{fig:XHfloors}}
\end{figure}

Because A{\sc eos} directly models Pop III enrichment, we can determine the median chemical abundances of gas in our Pop III halos and the scatter about that median. Figure \ref{fig:XHfloors} shows the [X/H] chemical abundances of gas in all pure Pop III halos at five different snapshots (equally spaced in time from about 150 Myr to about 300 Myr) with 16th to 84th percentile scatter. Even with increasing halo mass, the floors are fairly flat, though scatter is significant. We also find that the floors are also fairly flat with time, with the most scatter at early times (< 175 Myr). These floors depend on Pop III yields \citep{HegerWoosley2010} and critical metallicity of the transition from Pop III to Pop II stars (Z < 10$^{-5}$ Z$_\odot$).

As expected, most of the metal mass in the gas is oxygen, which is the most abundantly produced element in Pop III CCSNe (see Figure \ref{fig:yields}). We find a best-fit metallicity floor of about [O/H] = -4.0, which is lower than that assumed in EDGE, but they found that changing the floor to Z = 10$^{-4}$ Z$_\odot$ did not significantly affect their results \citep{Agertz2020}. For all of the best-fit metallicity floors, see the red lines in Figure \ref{fig:XHfloors}.

Our simulation supports a Pop III metallicity floor of Z = 10$^{-4}$ Z$_\odot$ that generally holds for halo masses of $10^{5}$ to $10^{7}$ M$_\odot$. At the highest halo masses, the metallicity of the gas increases, but there are only 4 halos above 5$\times 10^{6}$ M$_\odot$, so this may or may not be robust. The best-fit metallicity floor of each element is approximately: [C/H] = -4.1, [N/H] = -5.3, [O/H] = -4.0, [Na/H] = -4.9, [Mg/H] = -4.4, [Ca/H] = -4.9, [Mn/H] = -5.4, and [Fe/H] = -5.1. There is significant scatter, however, due to inhomogeneous mixing that follows localized enrichment from different SNe and explosive outflows. We exclude Sr and Ba as they are not produced in significant amounts by Pop III.

\begin{figure}[tb]
\center
\includegraphics[width=0.95\linewidth]{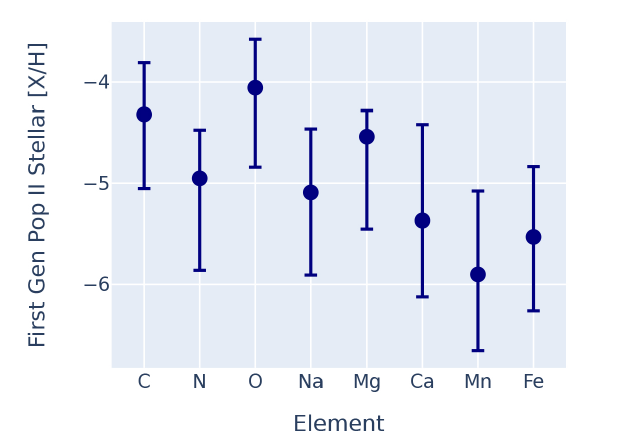}
\caption{The [X/H] stellar chemical abundances of the oldest low-mass Pop II stars in each of the 17 Pop II galaxies in the simulation, for each of eight metals (Sr and Ba are excluded because they exist in only trace amounts). Shown is the median across each of the oldest Pop II star formation events for each galaxy, with scatter representing 16th to 84th percentile.
\label{fig:popIIXH}}
\end{figure}

We also look at the stellar chemical abundances of the first low-mass Pop II stars in each galaxy to see how the gas enrichment translates into stellar enrichment for the first generation Pop II stars. 17 galaxies have begun Pop II star formation (see Figure \ref{fig:popIIgal}), and we plot the chemical abundances of the first-generation low-mass Pop II stars from each Pop II galaxy in Figure \ref{fig:popIIXH}. These abundances are similar to the gas metallicity floors in Figure \ref{fig:XHfloors}, as we would expect, with some differences (e.g., some of the median [X/H] abundances are noticeably lower) that are likely due to the smaller sample size and inhomogeneous metal mixing. As seen by the error bars, the scatter in the median chemical abundances is over a dex, showing significant variation in the abundances of the first generation of Pop II stars across different galaxies. The median [O/H] of the first generation of low-mass Pop II stars is [O/H] = -4.1, consistent with that of the Pop III halo gas.

\section{Conclusions}
\label{sec:conc}

In this paper, we introduce the A{\sc eos} project, a series of simulations designed to trace early galaxy formation processes and model the chemical enrichment of individual stars in unprecedented detail while including Pop III enrichment, radiative transfer, and galaxy formation in a cosmological context. Our simulations aim to shed light on the spread of observed stellar chemical abundance patterns in ultra-faint dwarf galaxies, metal mixing in the interstellar medium, the impact of Pop III stars, and more. We describe the methods and novel handling of star-by-star chemical enrichment.



By modeling individual stars with their chemical yields, we can capture the intricate interplay between galaxy evolution and different nucleosynthetic processes. As a proof of concept, we demonstrate that the detailed element tracking in the A{\sc eos} simulations allows us to reproduce scatter and structure in stellar chemical abundance space that can be directly linked to the progenitors of the stars. This star-by-star approach thus allows us to decode valuable information about hierarchical galaxy formation and early nucleosynthetic events.

We also explore the concept of how to treat small early galaxies: as individuals or as systems. We demonstrate examples of external enrichment in the simulation, where a more massive galaxy enriches the gas of nearby galaxies through outflows. The earliest galaxies live in low-mass halos that lose significant amounts of gas and metals, freely sharing with the galaxies near them. We consider these galaxies not to be individuals but to be systems that functionally evolve together.

Additionally, we estimate metallicity floors for different metals (e.g., [O/H] = -4) as a result of Pop III enrichment. These metallicity floors are fairly flat even with variations in halo mass and time, though with significant scatter. There is also over a dex of scatter in the median metallicity of first generation Pop II stars from different galaxies, showing that significant variation exists between first generation Pop II. We also investigate CEMP signatures in the first generation of Pop II stars and compare to the frequencies of CEMP stars observed in the Milky Way. [C/Fe] drops with metallicity in the simulations as it does in the observations.

The current simulations have been run from redshift $z=130$ to $z=14.5$, and future work will extend this work to the epoch of reionization for a suite of zoom-ins on a select sample of small galaxies that should be early analogs of the surviving ultra-faint dwarf galaxies. This will enable us to capture a comprehensive history of galaxy growth and different nucleosynthetic events that drove the associated processes. By exploring variations and inter-dependence of galaxy evolution and nucleosynthetic yields, we will quantify the origins of observed abundance scatter, the metal retention of small bursty galaxies, the impact of metal mixing in the early ISM.

\section*{Acknowledgments} K.B. is supported by an NSF Astronomy and Astrophysics Postdoctoral Fellowship under award AST-2303858. J.M. acknowledges support from the NSF Graduate Research Fellowship under grant DGE-2036197. J.H.W. acknowledges support by NSF grant AST-2108020 and NASA grants 80NSSC20K0520 and 80NSSC21K1053.  M.-M.M.L. and E.P.A were partly supported by NSF grants AST18-15461 and AST23-07950. E.P.A. also received support from NASA Astrophysical Theory grant 80NSSC24K0935.  G.L.B. acknowledges support from the NSF (AST-2108470 and AST-2307419, ACCESS), a NASA TCAN award, and the Simons Foundation through the Learning the Universe Collaboration.
A.F. acknowledges support from  NSF grant AST-2307436.

The authors acknowledge the Texas Advanced Computing Center at The University of Texas at Austin for providing HPC and storage resources that have contributed to the research results reported within this paper.



%
\appendix
\setcounter{section}{0}%
\renewcommand\thesection{\Alph{section}}
\counterwithin{figure}{section}

\section{Metal Yields}
\label{sec:appendix_yields}

\begin{figure}[b]
\center
\includegraphics[width=0.8\linewidth]{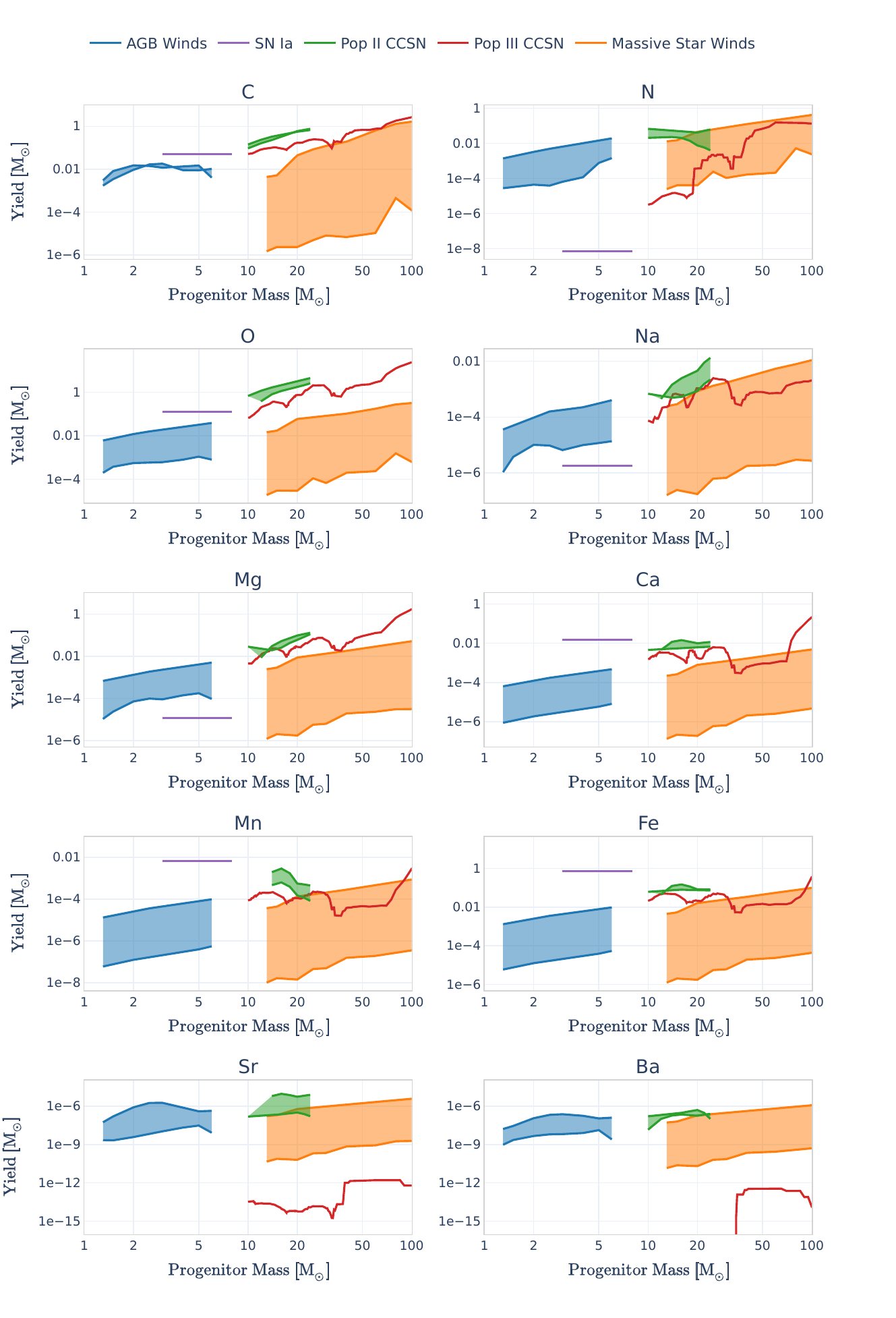}
\caption{The yields of each metal from each astrophysical source in the simulation. The shaded regions include the yield range for progenitor stars at different metallicities for a given progenitor mass. AGB wind yields are from \citet{Cristallo2015}, SN Ia yields are from \citet{Thielemann1986}, Pop II CCSN and massive star wind yields are from \citet{Limongi2018}, and Pop III yields are from \citet{HegerWoosley2010}. 
\label{fig:yields}}
\end{figure}

\begin{figure}[b]
\center
\includegraphics[width=0.8\linewidth]{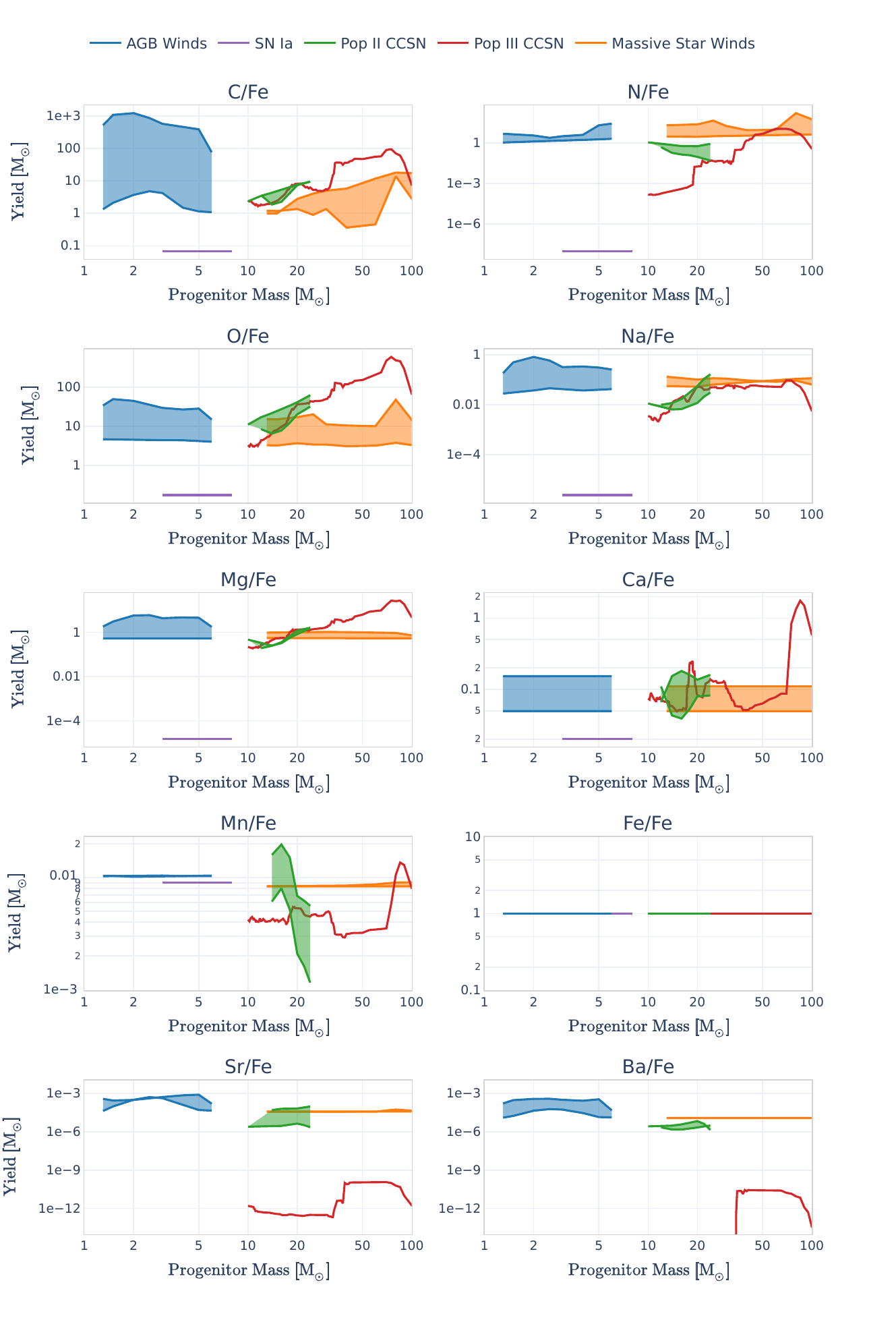}
\caption{The ratio of the yields of each metal from each astrophysical source in the simulation, with respect to iron (Fe). The shaded regions include the yield range for progenitor stars at different metallicities for a given progenitor mass. 
\label{fig:yieldsFe}}
\end{figure}

Figure \ref{fig:yields} illustrates the mass yields used in the A{\sc eos} simulations for ten tracked elements — C, N, O, Na, Mg, Ca, Mn, Fe, Sr, and Ba — across different stellar progenitor masses, highlighting contributions from various nucleosynthetic sources. These sources include Pop III core-collapse supernovae (CCSN), Pop II CCSN, AGB winds, SNIa, and massive star winds. Each curve represents the mass yield of an element as a function of progenitor mass, revealing the variations in yield patterns due to different evolutionary processes and initial conditions. The plot emphasizes the mass-dependent nature of element production, crucial for understanding the chemical evolution in our simulations. For more information, see Section \ref{sec:stellar yields}. We also show the ratio of the yields of each metal species with respect to Fe (Figure \ref{fig:yieldsFe}).

The yield sets used for this study were chosen based on a combination of factors, including alignment with observed data and ease of implementation in simulations. For AGB stars (M < 8 M$_\odot$), we selected the yields from \citet{Cristallo2015}, which were recommended for their ability to capture the production of elements like Sr and Ba. For massive stars (M > 8 M$_\odot$), we adopted the yields from \citet{Limongi2018} (LC18), as they offer a detailed treatment of stellar rotation and its impact on yields, despite some challenges like the underproduction of Mg. The LC18 yields also assume that stars above 25 M$_\odot$ undergo direct collapse without producing supernova yields, a feature consistent with previous assumptions in other models used by the team. This choice balances the need for physical realism with practical considerations of implementing yield tables that separate contributions from stellar winds and supernovae, allowing for a more flexible simulation setup. Despite the challenges of managing memory usage with many tracer fields, this combination of yield sets was found to produce the most realistic results when compared with Milky Way observations in galactic chemical evolution (GCE) models using OMEGA \citep{Cote18}, making it the preferred choice for the study.

\begin{figure}[tb]
\center
\includegraphics[width=0.7\linewidth]{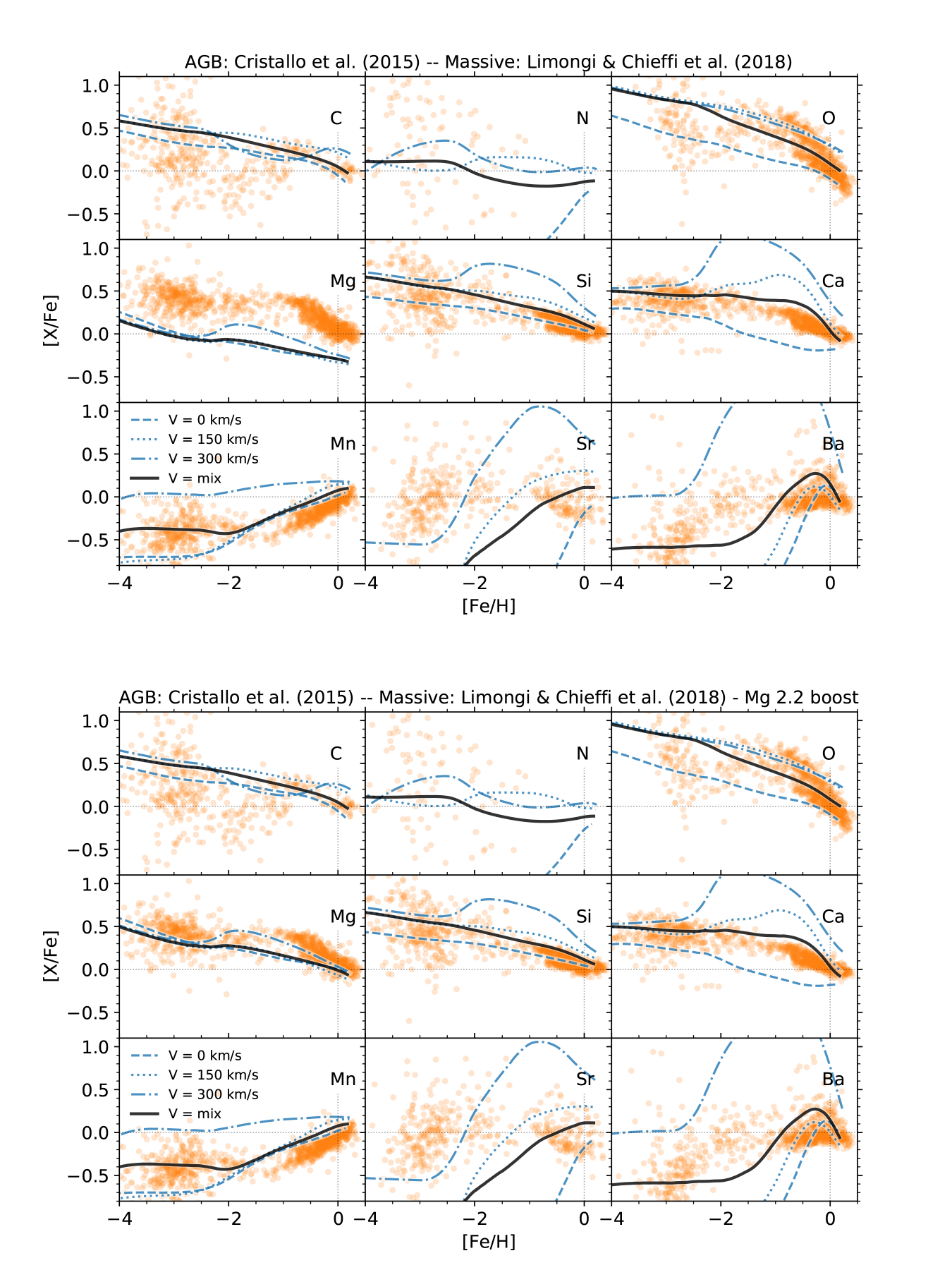}
\caption{\textit{Top:} Shown as lines, simulated stellar chemical abundances for different stellar rotation velocities in a galactic chemical evolution OMEGA model using AGB wind yields from \citet{Cristallo2015} and Pop II CCSN and massive star wind yields from \citet{Limongi2018}. In orange, observations of metal-poor stars in the Milky Way. \textit{Bottom:} Results after boosting the Mg yields from LC18 by a factor of 2.2 to match observations. 
\label{fig:Mgboost}}
\end{figure}

The decision to increase the magnesium (Mg) yields by a factor of 2.2 was made to resolve a mismatch between the theoretical yields predicted by the LC18 model and the observed [Mg/Fe] ratios in metal-poor stars of the Milky Way. The unmodified LC18 yields tended to underproduce Mg, leading to lower-than-expected [Mg/Fe] ratios in simulations that track the chemical evolution of galaxies. Observational data from stars with [Fe/H] < -2 indicate that Mg should be more abundant, reflecting the contributions of core-collapse supernovae in the early Universe. The 2.2x scaling factor was derived through iterative comparisons between simulated results using our chosen yields with OMEGA \citep{Cote18} and stellar chemical abundances of metal-poor stars in the Milky Way from JINAbase \citep{JINAbase}, effectively bringing the simulated [Mg/Fe] ratios into alignment with what is observed (see Figure \ref{fig:Mgboost}). This increase in Mg does not affect gas cooling in our simulation because it does not provide significant cooling \citep[e.g.,][]{GrackleMethod} and in \Aeos the gas cooling is determined by total metallicity. While the Mg adjustment is empirical and does not address the physical processes behind the underproduction, it ensures that the simulations more accurately reflect the chemical enrichment history of early stellar populations, improving the overall consistency of the model with observed data.

 \clearpage
\bibliographystyle{yahapj}
\bibliography{refs}

\end{document}